\documentclass[aps,pra,twocolumn,amsmath,amssymb,nofootinbib,superscriptaddress]{revtex4-1}

\usepackage{graphicx}
\usepackage{mathrsfs}
\usepackage[usenames,dvipsnames]{color}
\usepackage{hyperref}
\hypersetup{colorlinks = true, linkcolor = [rgb]{0.19411,0.51882,0.667058}, urlcolor = [rgb]{0.125490,0.29542,0.1647058}, citecolor = [rgb]{0.75882,0.37411,0.14117}}
\usepackage{comment}
\usepackage{amsmath, amssymb}
\usepackage{mathtools}
\usepackage{notes2bib}
\usepackage{float}
\usepackage[normalem]{ulem}
\usepackage{soul}
\usepackage{appendix}
\usepackage{bm}
\usepackage{cleveref}
\usepackage{float}
\setstcolor{red}

\begin{document}

\bibliographystyle{apsrev}

\title{Sub-shot-noise-limited phase estimation via SU(1,1) interferometer with thermal states}

\author{Xiaoping Ma}
\email[]{cherry901115@hotmail.com}
\affiliation{Department of Physics, Ocean University of China, Qingdao, 266100, China}
\affiliation{Hearne Institute for Theoretical Physics and Department of Physics \& Astronomy, Louisiana State University, Baton Rouge, LA 70803, United States}

\author{Chenglong You}
\affiliation{Hearne Institute for Theoretical Physics and Department of Physics \& Astronomy, Louisiana State University, Baton Rouge, LA 70803, United States}

\author{Sushovit Adhikari}
\affiliation{Hearne Institute for Theoretical Physics and Department of Physics \& Astronomy, Louisiana State University, Baton Rouge, LA 70803, United States}

\author{Elisha S. Matekole}
\affiliation{Hearne Institute for Theoretical Physics and Department of Physics \& Astronomy, Louisiana State University, Baton Rouge, LA 70803, United States}

\author{Hwang Lee}
\affiliation{Hearne Institute for Theoretical Physics and Department of Physics \& Astronomy, Louisiana State University, Baton Rouge, LA 70803, United States}

\author{Jonathan P. Dowling}
\affiliation{Hearne Institute for Theoretical Physics and Department of Physics \& Astronomy, Louisiana State University, Baton Rouge, LA 70803, United States}

\date{\today}

\begin{abstract}
We theoretically study the phase sensitivity of an SU(1,1) interferometer with a thermal state and squeezed vacuum state as inputs and parity detection as measurement. We find that phase sensitivity can beat the shot-noise limit and approaches the Heisenberg limit with increasing input photon number.
\end{abstract}

\maketitle

\section{Introduction}

Over the past decades, there has been much progress in both theoretical and experimental research on quantum metrology. This development has been due to the use of quantum resources in quantum metrology and has led us to surpass  classical sensitivity\ \cite{Michelson1887,Xiao1987,Barish1999,Lee2002,Boto2000,Escher2012,Giovannetti2004,Giovannetti2011,Schnabel2010,Ou2012,You17,Gard17}. One of the important tasks in quantum metrology is the estimation of optical phase. Mach-Zehnder interferometers (MZI), which consist of two mirrors, two beam splitters and a phase shifter, is a conventional tool for phase estimation. Using only classical light, MZI can achieve a phase sensitivity of ${1 / {\sqrt N }}$, called shot noise limit (SNL) , where $N$ is the mean number of photons that have passed through the MZI. But using quantum resources, the phase sensitivity can approach $1/{N}$ which is called Heisenberg limit\ \cite{Caves1981,Bollinger1996,Dowling2008}. A MZI is also called an SU(2) interferometer as Yurke \emph{et al}. showed that the group SU(2) can naturally characterize MZI in \cite{Yurke1986}. Furthermore, in the same paper, the authors first proposed another type of interferometer, which is characterized by the group SU(1,1). They named it SU(1,1) interferometer and showed that it can beat SNL. SU(1,1) interferometer is similar to SU(2) interferometer except that the two beam splitters are replaced by two optical parametric amplifiers (OPAs).

Some new SU(1,1) interferometer schemes were proposed recently. Plick \emph{et al}.\ \cite{Plick2010} presented a theoretical scheme that uses bright coherent source to boost the sensitivity of SU(1,1) interferometer. Their scheme achieved scaling far below SNL for bright sources. To get the same sensitivity and reduce the required intensities of the input states, Li \emph{et al}.\ \cite{Li2014} considered a squeezed vacuum state to replace one of the two input coherent states. In addition, they used homodyne measurement as their detection, which is convenient for experimental detection of squeezing.  Li \emph{et al}.\ \cite{Li2016} also proposed another scheme that uses parity detection, which simply measures the even or odd number of photons in the output mode, and can be implemented by using homodyne techniques\ \cite{Anisimov2010}. More recently, Szigeti \emph{et al}.\ \cite{Szigeti2017} presented a modification of a SU(1,1) interferometer where all the input particles participate in phase measurement and showed how this can be implemented in spinor Bose-Einstein condensates and hybrid atom-light systems.

All SU(1,1) interferometer schemes mentioned above estimates phase information with coherent states. In this article, we propose an interferometric scheme that performs phase estimation using a thermal state. Thermal states are central concept in thermodynamics and statistical mechanics and have several important properties, such as, typical dynamics evolve the system towards thermal state, and it is the state that maximizes the entropy under constraints on physical quantities\ \cite{Jaynes1957,Nicole2016}. Moreover, a thermal state is more easily accessible and, being an equilibrium state, is more stable than coherent state. This makes our SU(1,1) interferometric scheme much easier to implement experimentally. 

The structure of the present paper is as follows: In Sec.\ \uppercase\expandafter{\romannumeral2.\ A}, we introduce the model along with basic transformation rules of SU(1,1) interferometer. The Heisenberg limit and shot-noise limit of the system is calculated in Sec.\ \uppercase\expandafter{\romannumeral2.\ B}. The Parity detection method along with phase sensitivity obtained from this scheme is presented in Sec.\ \uppercase\expandafter{\romannumeral2.\ C}. Finally, comparison between phase sensitivity and Heisenberg limit is given in Sec.\ \uppercase\expandafter{\romannumeral2.\ D}. Lastly,  we conclude with a brief summary in Sec.\ \uppercase\expandafter{\romannumeral3}.

\section{SU(1,1) INTERFEROMETER VIA PARITY DETECTION}
\subsection{Model}
\begin{figure}
\centering
\includegraphics[width=0.48\textwidth]{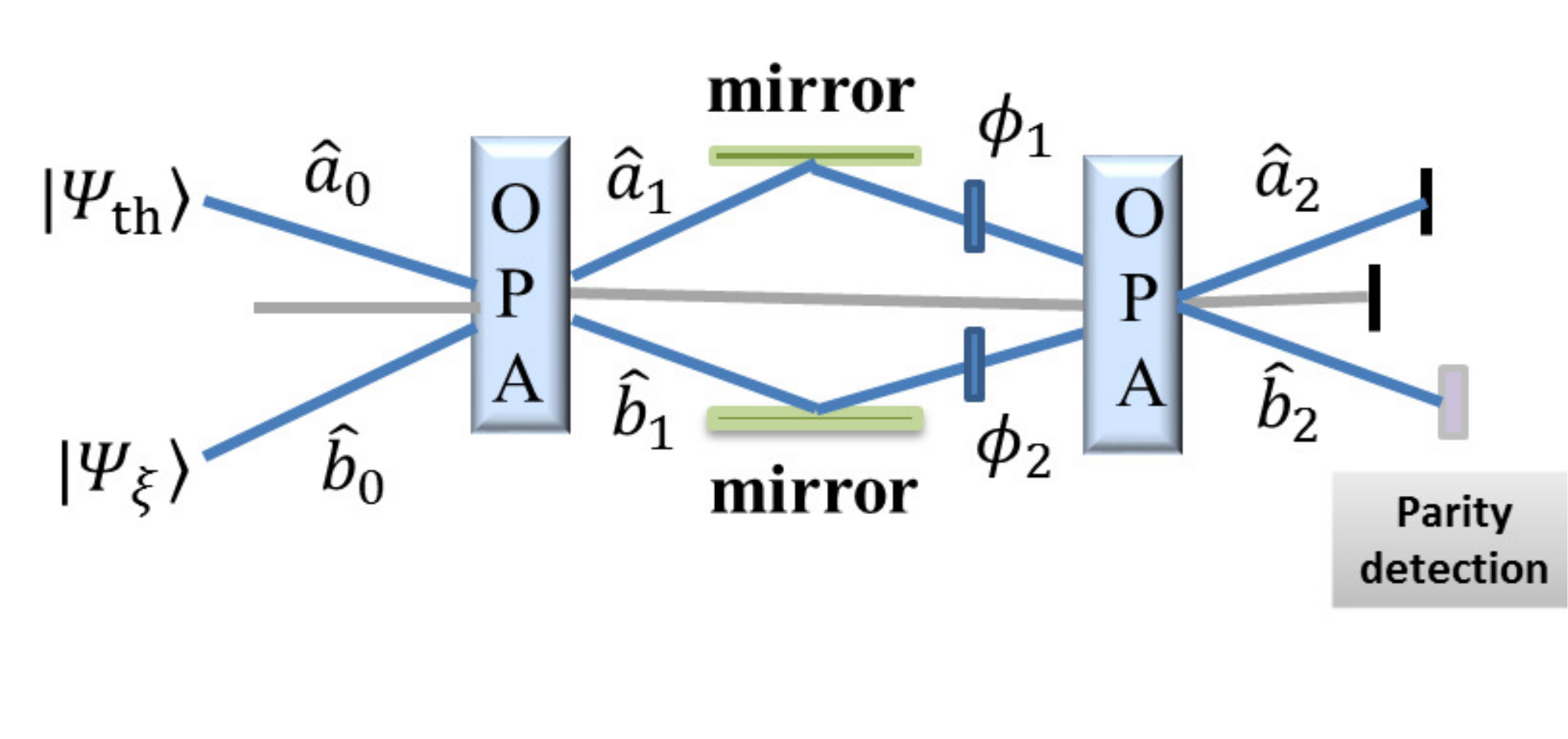}
\caption{\label{fig1}Schematic diagram of an ideal SU(1,1) interferometer. An SU(1,1) interferometer is similar to a MZI except the two beam splitters are replaced by two OPAs. $\left| {{\Psi _{\textrm{th}}}} \right\rangle $ and $\left| {{\Psi _\xi }} \right\rangle $ are thermal state and squeezed vacuum state respectively. ${a_i}$ and ${b_i} (i=0,1,2)$ denote two light beams in the different processes. $\phi_{1} $ and $\phi_{2} $ describe the phase shift in the two paths.}
\end{figure}

An SU(1,1) interferometer is shown in Fig.\ \ref{fig1}, in which beam splitters of a traditional MZI are replaced by OPAs. Here, we consider a thermal state and a squeezed vacuum state as inputs. Let $\hat a$ $({\hat a^\dag })$, $\hat b$ $({\hat b^\dag })$ be the annihilation (creation) operators of two modes respectively. After first OPA, the upper path undergoes a phase shift ${\phi _{\rm{1}}}$ and the lower path undergoes a phase shift ${\phi _{\rm{2}}}$. After second OPA, we perform output detection to obtain the acquired phase difference between the two modes. 

The evolution through an SU(1,1) interferometer is as follows. Transformations through first OPA, phase shifter and second OPA are described by

\begin{equation}\label{eq1}
{\hat T_{\text{OPA1}}} = \left( {\begin{array}{*{20}{c}}
{{u_1}}&{{v_1}}\\
{v_1^*}&{{u_1}}
\end{array}} \right),
\end{equation}

\begin{equation}\label{eq2}
  {\hat T_{\phi}} = \left( {\begin{array}{*{20}{c}}
  {{e^{i{\phi _{\rm{1}}}}}}&0\\
  0&{{e^{i{\phi _{\rm{2}}}}}}
  \end{array}} \right),
\end{equation}

\begin{equation}\label{eq3}
{\hat T_{\text{OPA2}}} = \left( {\begin{array}{*{20}{c}}
{{u_2}}&{{v_2}}\\
{v_2^*}&{{u_2}}
\end{array}} \right),
\end{equation}
with ${u_j} = \cosh{g_j} $, ${v_j} = {e^{i{\theta _j}}}\sinh{g_j}$, where ${\theta _j}$ and ${g_j}$ are phase shift and parametrical strength in the $\text{OPAs}$ $\left( {j = 1,2} \right)$. Hence, the tranformation through a SU(1,1) interferometer is represented by $\hat T = {\hat T_{\text{OPA2}}}{\hat T_\phi }{\hat T_{\text{OPA1}}}$.

After the first OPA, the relation between modes ${\hat a_1},{\hat b_1}$ and ${\hat a_0},{\hat b_0}$ is described by
\begin{equation}\label{eq5}
\left( {\begin{array}{*{20}{c}}
{{{\hat a}_1}}\\
{\hat b_1^\dag }
\end{array}} \right) = \hat T_{\text{OPA1}}\left( {\begin{array}{*{20}{c}}
{{{\hat a}_0}}\\
{\hat b_0^\dag }
\end{array}} \right).
\end{equation}

After propagation through SU(1,1) interferometer, the relation between input and output modes is described by
\begin{equation}\label{eq4}
\left( {\begin{array}{*{20}{c}}
{{{\hat a}_2}}\\
{\hat b_2^\dag }
\end{array}} \right) = \hat T\left( {\begin{array}{*{20}{c}}
{{{\hat a}_0}}\\
{\hat b_0^\dag }
\end{array}} \right).
\end{equation}

\subsection{Heisenberg limit and Shot-noise limit}

From Eq.\ (\ref{eq5}), we can calculate the mean photon number $\bar n$ inside the SU(1,1) interferometer. Notice that $\bar n$ is not the total input photon number because OPAs are nonlinear and have gain and there will be spontaneous photons emitted. Hence, the photon number inside SU(1,1) interferometer is amplified compared to input number. The mean photon number $\bar n$ inside SU(1,1) interferometer is given by
\begin{equation}\label{eq6}
\bar n = \left\langle {{\Psi _\text{in}}} \right|( {\hat a_1^\dag {{\hat a}_1} + \hat b_1^\dag {{\hat b}_1}} )\left| {{\Psi _\text{in}}} \right\rangle, \end{equation}
with $\left| {{\Psi _\text{in}}} \right\rangle  = \left| {{\Psi _\text{th}}} \right\rangle  \otimes \left| {{\Psi _\xi }} \right\rangle $, where $\left| {{\Psi _\text{th}}} \right\rangle $ is a thermal state and $\left| {{\Psi _\xi }} \right\rangle $ is a squeezed vacuum state.
Finally, the total photon number in SU(1,1) interferometer is given by 
\begin{equation}
\bar n = {{{( {{n_\text{OPA}} + 1} )}}\left( {{n_\text{th}} + {n_\text{s}} } \right) + {n_\text{OPA}} },
\end{equation}
where, ${{n_\text{th}}}$ is the input photon number of the thermal state, ${{n_\text{s}} = \sinh{^2}r}$ is the photon number of the squeezed vacuum, $n_\text{OPA}=2{\sinh ^2}g$ is the emitted photon number from the first OPA and $g=g_1=g_2$ is the parametrical strength of the two OPAs. We can see that there are two contributions in increasing the mean photon number inside SU(1,1) interferometer. The first contribution comes from the amplification process of the input photon number and the second contribution appears due to the amplification of the  input vacuum state, a process called the spontaneous process. Using the total inside photon number, for our SU(1,1) scheme with thermal and squeezed vacuum state, we calculate the SNL and HL to be
\begin{equation}\label{eq7}
\Delta {\phi _\text{SNL}} = \frac{1}{{\sqrt {\bar n} }}{\rm{ = }}\frac{{\rm{1}}}{\sqrt {{{{( {{n_\text{OPA}} + 1} )}}\left( {{n_\text{th}} + {n_\text{s}} } \right) + {n_\text{OPA}} }}} ,
\end{equation}
\begin{equation}\label{eq8}
\Delta {\phi _\text{HL}} = \frac{1}{{\bar n}}{\rm{ = }}\frac{{\rm{1}}}{{{( {{n_\text{OPA}} + 1} )}}\left( {{n_\text{th}} + {n_\text{s}} } \right) + {n_\text{OPA}} }.
\end{equation}

\subsection{Phase sensitivity}

We consider parity measurement as our output detection. Parity detection is a single mode measurement and parity operator detection on output mode \emph{b} is defined as 
\begin{equation}\label{eq9}
{{\hat \Pi }_b}  \equiv {\left( { - 1} \right)^{\hat b_2^\dag {{\hat b}_2}}}.
\end{equation}
According to Ref.\ \cite {Anisimov2010}, parity measurement satisfies $\langle {{{\hat \Pi }_b}}\rangle  = \pi W\left( {0,0} \right)$, that is, the expectation of parity measurement is given by the value of Wigner function at the origin of the phase space. This property is very useful for calculating the parity signal. 
The minimum detectable phase i.e., the phase sensitivity is given by
\begin{equation}\label{eq10}
\Delta \phi  = \frac{  {{{\Delta \hat \Pi }_b}} } {{\left| {{{\partial \langle {{{\hat \Pi }_b}} \rangle }}/{{\partial \phi }}} \right|}},
\end{equation}
where, $ {{{\Delta \hat \Pi }_b}}   = \sqrt {\langle {\hat \Pi _b^2} \rangle  - {{\langle {{{\hat \Pi }_b}} \rangle }^2}}  = \sqrt {1 - {{\langle {{{\hat \Pi }_b}} \rangle }^2}} $ with the property that ${\langle {\hat \Pi _b^2} \rangle  = 1}$.
The phase sensitivity via parity detection for a SU(1,1) interferometer with thermal and squeezed vacuum state is calculated to be
\begin{equation}\label{eq11}
\Delta \phi  = \sqrt {\frac{2}{{{n_\text{OPA}}( {{n_\text{OPA}} + 2} )[ {1 +( {1 + 2{n_s}} )( {1 + 2{n_\text{th}}} )} ]}}} ,
\end{equation}
 with the assumption, $\phi {\rm{ = 0}}$. For an arbitrary value of $\phi$, the phase fluctuation is shown in \cref{eq48,eq49} (See Appendix B).

\subsection{COMPARISON BETWEEN PHASE SENSITIVITY AND HL}

In this section, we compare the phase sensitivity via parity detection with HL and SNL.

\begin{figure}
\centering
\includegraphics[width=0.4\textwidth]{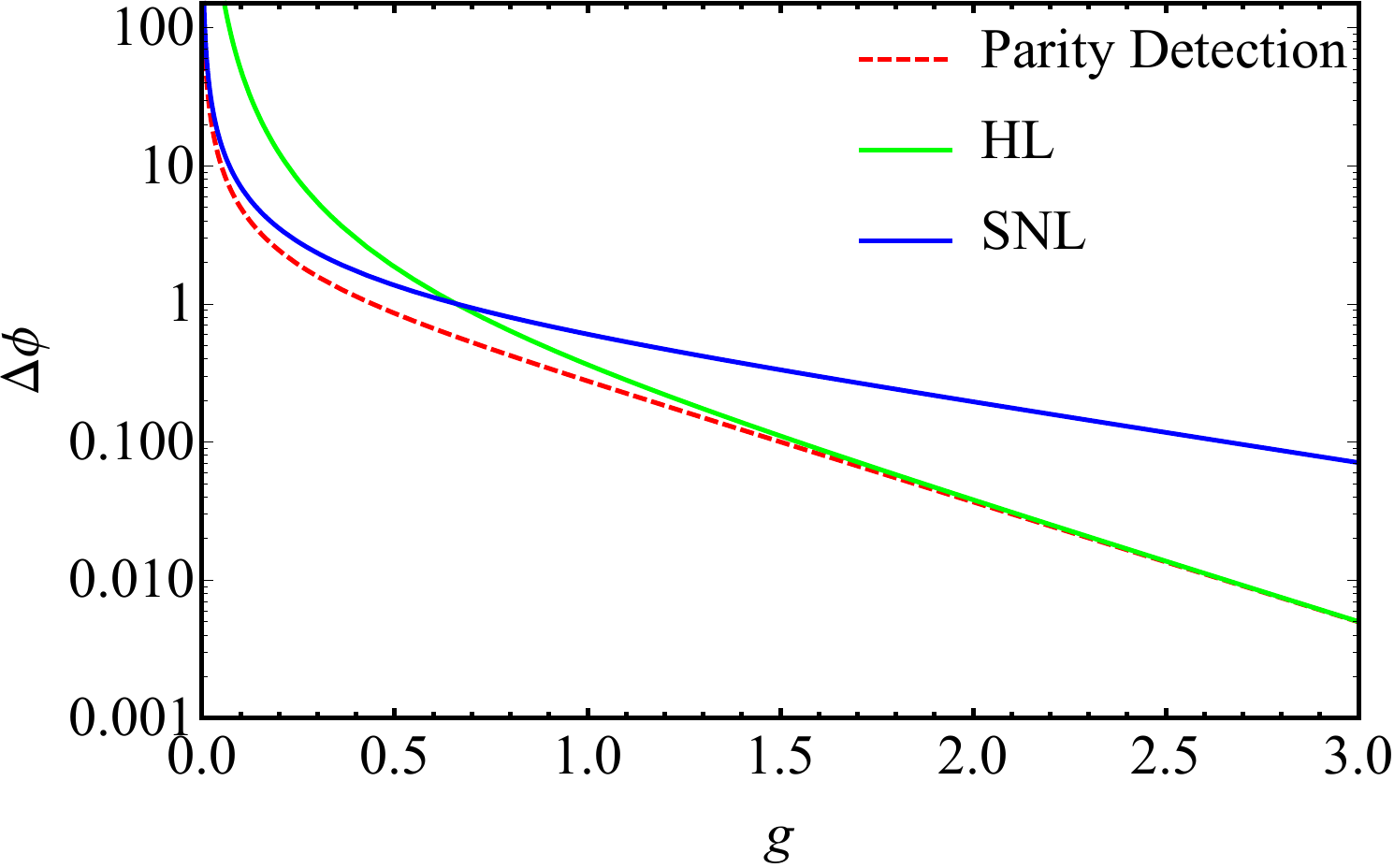}
\caption{\label{fig2} Phase sensitivity with parity detection as a function of $g$ with constraint $r = 0$, $n_{\text{th}}=0$ on the input.}
\end{figure}

First, we consider two vacuum states as our inputs. The phase sensitivity with parity detection is found to be $\Delta {\phi _v} =1/\sqrt{{n_\text{OPA}}( {{n_\text{OPA}} + 2} )}$, while the corresponding Heisenberg limit is $\Delta {\phi _\text{HL}} =1/{n_\text{OPA}}$, and the shot noise limit is $\Delta {\phi _\text{SNL}} =1/\sqrt{{n_\text{OPA}}}$. We plot the phase sensitivity $\Delta \phi $ as a function of $g$ in Fig.\ \ref{fig2} along with $\Delta {\phi _\text{SNL}}$ and $\Delta {\phi _\text{{HL}}}$. In this case, phase sensitivity with parity detection always beats SNL and HL. Moreover, when $g \le 0.6$, SNL is below HL because the total inside photon number, ${n_{\text{OPA}}} < 1$.

\begin{figure}
\centering
\includegraphics[width=0.4\textwidth]{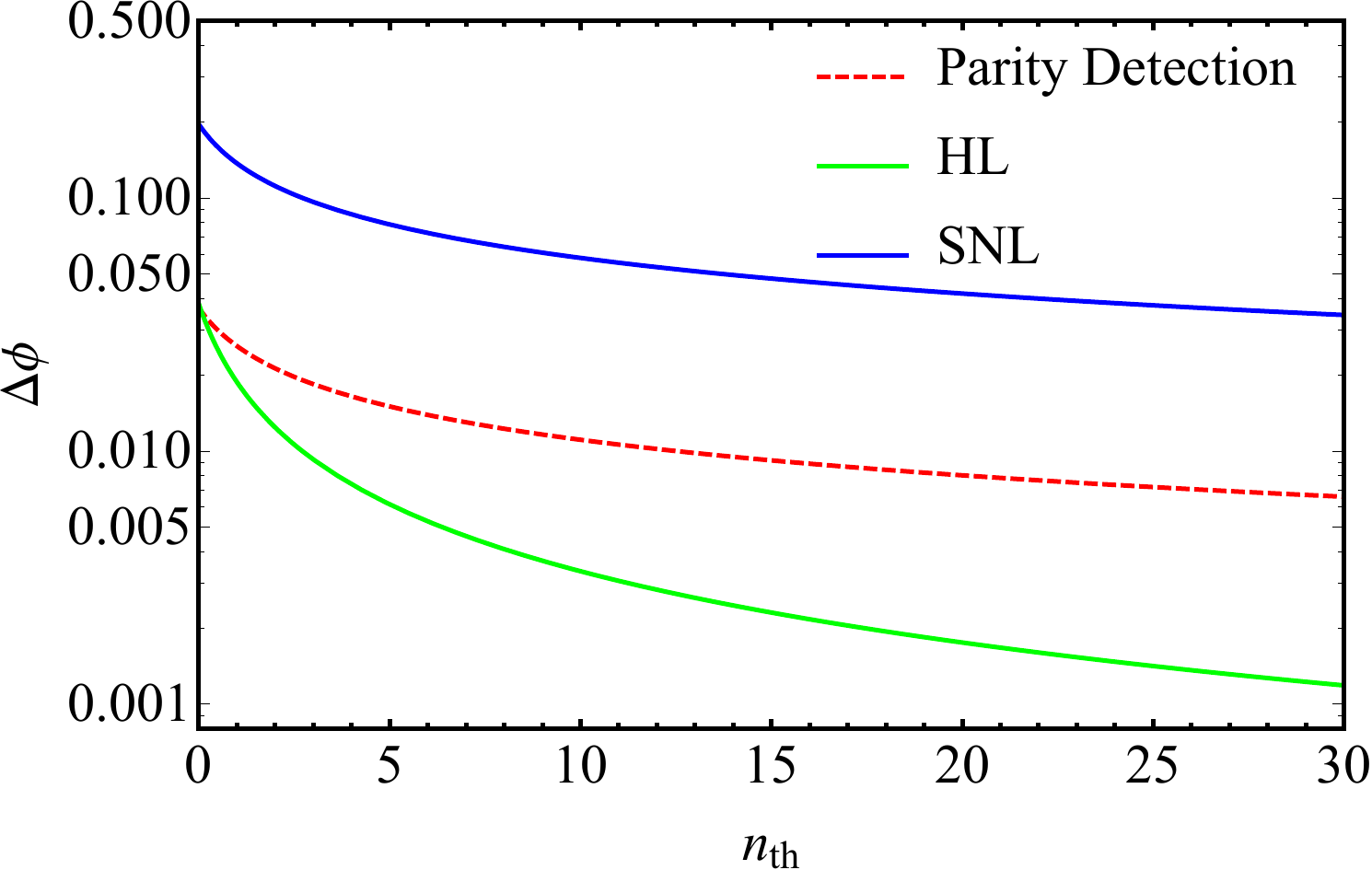}
\caption{\label{fig3} Phase sensitivity with parity detection as a function of ${n_{\text{th}}}$ with $r = 0$, $g = 2$. The phase sensitivity at ${n_{\text{th}}}=0$ is equal to the sensitivity at $g = 2$ in Fig.\ \ref{fig2} and becomes greater with increasing ${n_{\text{th}}}$.}
\end{figure}

We then, consider thermal and vacuum state as inputs. As depicted in  Fig.\ \ref{fig3}, under the condition ($r=0$ and $g=2$), phase sensitivity via parity detection always beat SNL but it does not approach HL. In this case, as the total inside photon number is always larger than 1, SNL is always greater than HL. Note that the values of phase sensitivity and HL at ${n_{\text{th}}}=0$ are equal to the corresponding values at $g=2$ in Fig.\ \ref{fig2}. The phase sensitivity increases by increasing the mean photon number of the thermal state.

\begin{figure}
\centering
\includegraphics[width=0.4\textwidth]{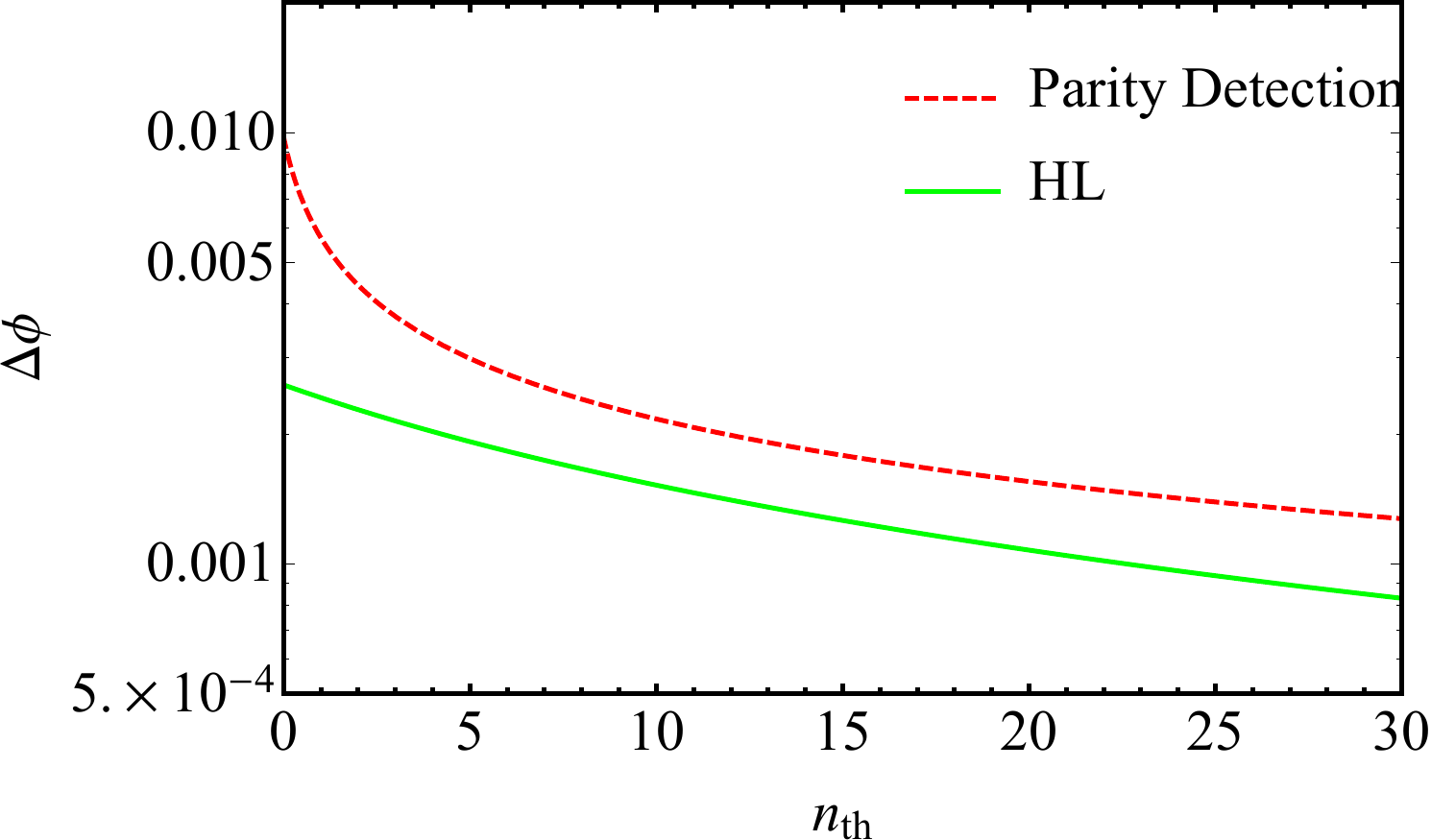}
\caption{\label{fig4} Phase sensitivity with parity detection as a function of ${n_{\text{th}}}$ with constraints $r = 2$, $g = 2$. Comparing the values at $n_{\text{th}}=20$ in Fig.\ \ref{fig3}, both the phase sensitivity and HL are lower due to replacing vacuum state with squeezed vacuum state. Moreover, the phase sensitivity is much higher.}
\end{figure}

Lastly, we inject thermal and squeezed vacuum state as inputs. Comparing Eq.\ (\ref{eq8}) with Eq.\ (\ref{eq11}), the necessary optimal condition for approaching HL is found to be
\vspace{-1.5mm}
\begin{equation}\label{eq12}
{n_\text{th}} = \frac{{\sinh }^2g - {n_s}} {{\cosh }^2\left( {2g} \right)}.
\end{equation}
The above expression guarantees that the phase sensitivity for input thermal state ${n_\text{th}}$, input squeezed vacuum state ${n_s}$ and the OPA process $g$ approaches HL. One can prove that regardless the values of ${n_s}$ and $g$, the optimal mean  photon number ${n_\text{th}}$ of the thermal state is no more than 1. 
Fig.\ \ref{fig4} shows phase sensitivity as a function of ${n_{\textrm{th}}}$ under the condition $r = 2$, $g = 2$. The phase sensitivity with parity detection is always below SNL but cannot beat HL. Comparing Fig.\ \ref{fig4} with Fig.\ \ref{fig3}, we see that the phase sensitivity becomes larger and closer to HL by replacing vacuum state with squeezed vacuum state. With increasing ${n_{\textrm{th}}}$, the phase sensitivity approaches the HL.

\begin{figure}
\centering
\includegraphics[width=0.4\textwidth]{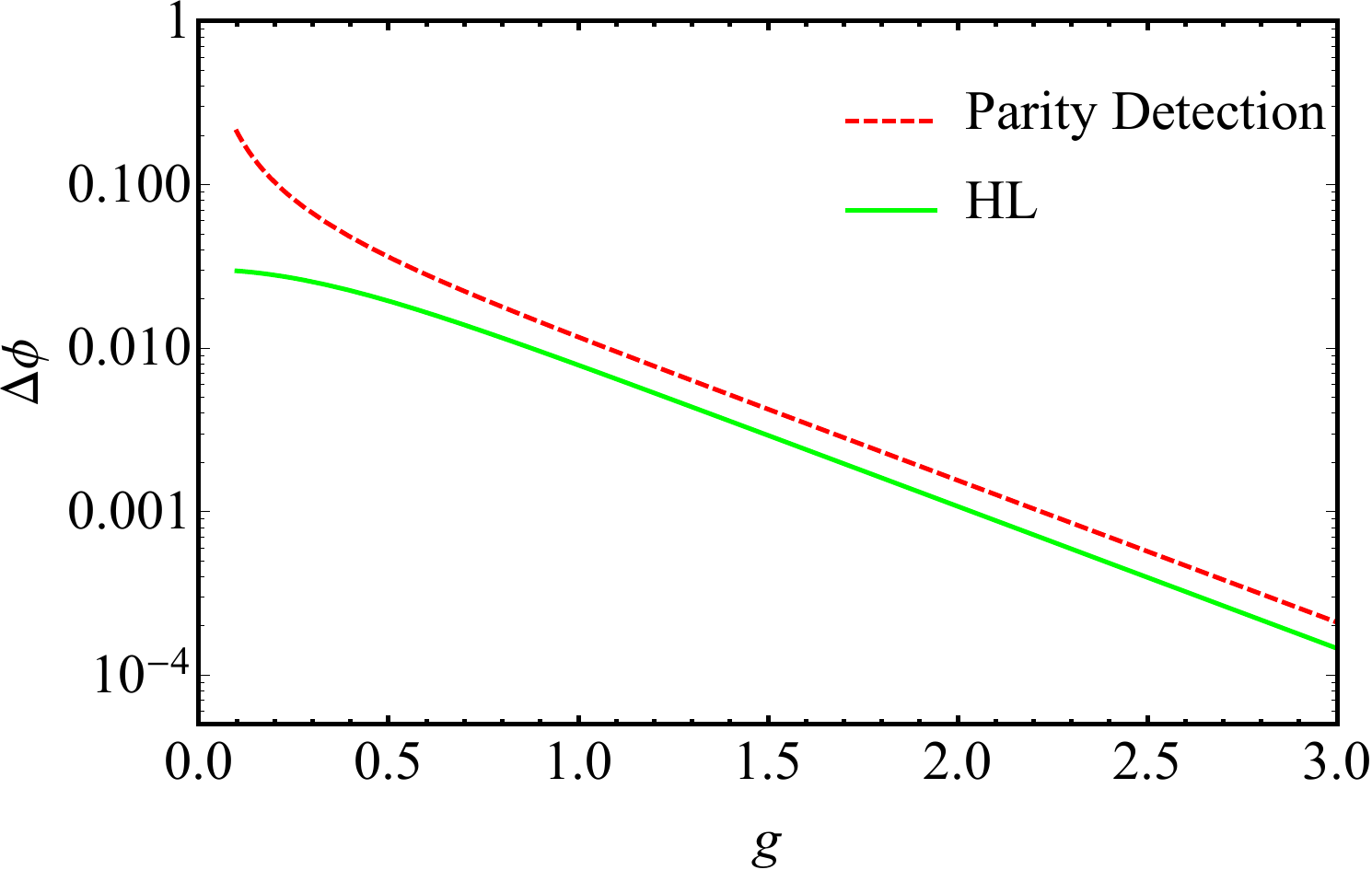}
\caption{\label{fig5} Phase sensitivity with parity detection as a function of $g$ with $r = 2$ and ${n_{\text{th}}} = 20$. The values of phase sensitivity and HL at $g=2$ are equal to the values at ${n_{\text{th}}} = 20$ in Fig.\ \ref{fig3} and increases with increase in $g$. }
\end{figure}

\begin{figure}
\centering
\includegraphics[width=0.4\textwidth]{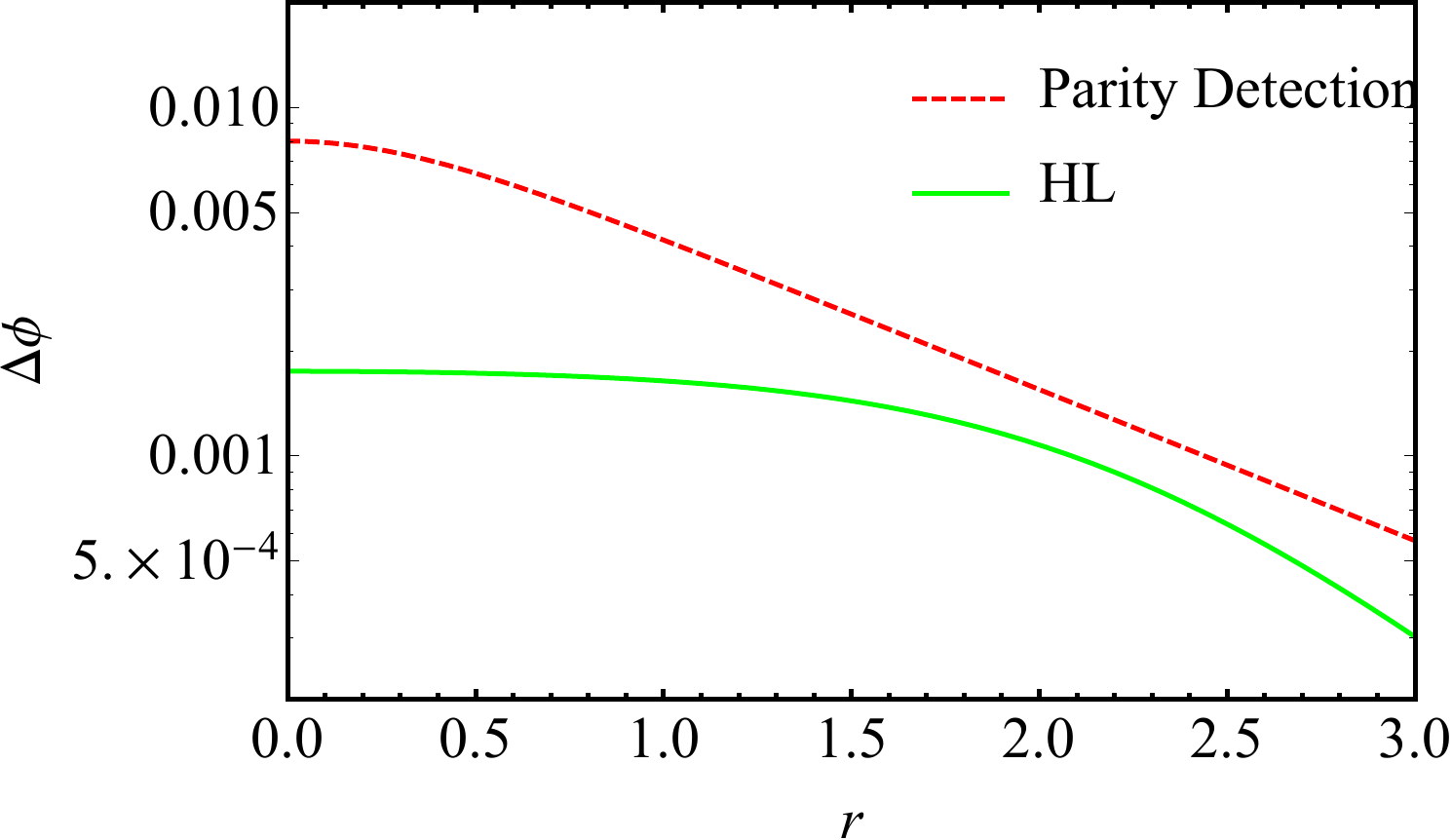}
\caption{\label{fig6} Phase sensitivity with parity detection as a function of $r$ under the condition $g = 2$ and ${n_{\text{th}}} = 20$. The values of phase sensitivity and HL at $g=2$ are equal to the values at ${n_{\text{th}}} = 20$ in Fig.\ \ref{fig3} and gets smaller by increasing pump power.}
\end{figure}

In Fig.\ \ref{fig5}, we compare phase sensitivity of parity detection with HL as a function of $g$. One can see that phase sensitivity at $g=2$ in Fig.\ \ref{fig5} is equal to the phase sensitivity at $n_\text{th}=20$ in Fig.\ \ref{fig3} and becomes closer to HL with increasing $g$. We also plot phase sensitivity and HL as a function of $r$ in Fig.\ \ref{fig6}. Similarly, the phase sensitivity at $r=2$ in Fig.\ \ref{fig6} is equal to the phase sensitivity at $n_\text{th}=20$ in Fig.\ \ref{fig3} and get closer to the HL with the increase in $r$.

\section{Conclusion}
We have studied the phase sensitivity of an SU(1,1) interferometer. If two vacuum states are considered at the input, the phase sensitivity can approach HL due to the total photon number inside the interferometer being very small. However, if one input state is a thermal state, the phase sensitivity worsens and at best beats the SNL but not the HL. If we replace the vacuum state by a squeezed vacuum state, different strategies will give higher phase sensitivity. These strategies include, increasing photon number of the thermal state or increasing the parametric strength g in the OPA or increasing the squeezed strength $r$ which all achieve higher phase sensitivity and approaches HL.

\section*{ACKNOWLEDGEMENT}
XPM acknowledges financial support from National Natural Science Foundation of China (Grant No. 61575180, 61701464, 11475160, 61640009), the Natural Science Foundation of Shandong Province (Grants No. ZR2014AQ026 and No.ZR2014AM023) and CSC. CY would like to acknowledge support from Economic Development Assistantship from Louisiana State University System Board of Regents. SA, ESM, HL and JPD would like to acknowledge support from Air Force Office of Scientific Research, Army Research Office, the Defense Advanced Research Project Agency, National Science Foundation, and Northrop Grumman Corporation.

\appendix
 \section{ PARITY DETECTION SIGNAL}

For analysis of our interferometer, we use Wigner functions as our mathematical model. First, we focus on evolution of mean values and covariance matrix of quadrature operators in SU(1,1) interferometer. The quadrature
operators can be defined as\ \cite{Braunstein2005}
\begin{equation}\label{eq15}
{{\hat x}_{{a_j}}} = {{\hat a}_j} + \hat a_j^\dag, \ {{\hat p}_{{a_j}}} =  - i({{\hat a}_j} - \hat a_j^\dag ),  \tag{A1}
\end{equation}
\begin{equation}\label{eq16}
{{\hat x}_{{b_j}}} = {{\hat b}_j} + \hat b_j^\dag,\ {{\hat p}_{{a_j}}} =  - i({{\hat b}_j} - \hat b_j^\dag),\tag{A2}
\end{equation}
and the quadrature column vector is defined as
\begin{equation}\label{eq17}
\bm{{X_j} }= ( {{{\hat X}_{j,1}},{{\hat X}_{j,2}},{{\hat X}_{j,3}},{{\hat X}_{j,4}}} )^T = ( {{{\hat x}_{{a_j}}},{{\hat p}_{{a_j}}},{{\hat x}_{{b_j}}},{{\hat p}_{{b_j}}}} )^T.\tag{A3}
\end{equation}
The mean and covariances of quadrature operators are given by
\begin{equation}\label{eq18}
\bm{{\bar X}_j} = {( {\langle {{{\hat X}_{j,1}}} \rangle ,\langle {{{\hat X}_{j,2}}} \rangle ,\langle {{{\hat X}_{j,3}}} \rangle ,\langle {{{\hat X}_{j,4}}} \rangle } )^T},\tag{A4}
\end{equation}
\begin{equation}\label{eq19}
\Gamma _j^{kl} = \frac{1}{2}\text{Tr}[ {( {{{\tilde X}_{j,k}}{{\tilde X}_{j,l}} + {{\tilde X}_{j,l}}{{\tilde X}_{j,k}}} )\rho } ],\tag{A5}
\end{equation}
where, ${{\tilde X}_{j,k}} = {{\hat X}_{j,k}} - \langle {{{\hat X}_{j,k}}} \rangle$, ${{\tilde X}_{j,l}} = {{\hat X}_{j,l}} -\langle {{{\hat X}_{j,l}}} \rangle$ and $\rho $ is a density matrix of the input state.

Transformations through the first OPA, phase shifter and the second OPA in  phase space are described by
\begin{equation}\label{eq20}
{S_\text{OPA1}} = \left( {\begin{array}{*{20}{c}}
{\begin{array}{*{20}{c}}
{\cosh g}\\
0\\
{  \sinh g}\\
0
\end{array}}&{\begin{array}{*{20}{c}}
0\\
{\cosh g}\\
0\\
{-\sinh g}
\end{array}}&{\begin{array}{*{20}{c}}
{ \sinh g}\\
0\\
{\cosh g}\\
0
\end{array}}&{\begin{array}{*{20}{c}}
0\\
{-\sinh g}\\
0\\
{\cosh g}
\end{array}}
\end{array}} \right),\tag{A6}
\end{equation}
\begin{equation}\label{eq21}
{S_\phi } = \left( {\begin{array}{*{20}{c}}
{\begin{array}{*{20}{c}}
{\cos ( {\frac{\phi }{2}})}\\
{\sin ( {\frac{\phi }{2}} )}\\
0\\
0
\end{array}}&{\begin{array}{*{20}{c}}
{ - \sin ( {\frac{\phi }{2}} )}\\
{\cos ( {\frac{\phi }{2}} )}\\
0\\
0
\end{array}}&{\begin{array}{*{20}{c}}
0\\
0\\
{\cos ( {\frac{\phi }{2}} )}\\
{\sin ( {\frac{\phi }{2}} )}
\end{array}}&{\begin{array}{*{20}{c}}
0\\
0\\
{ - \sin ( {\frac{\phi }{2}})}\\
{\cos( {\frac{\phi }{2}} )}
\end{array}}
\end{array}} \right),\tag{A7}
\end{equation}
\begin{equation}\label{eq22}
{S_\text{OPA2}} = \left( {\begin{array}{*{20}{c}}
{\begin{array}{*{20}{c}}
{\cosh g}\\
0\\
{ - \sinh g}\\
0
\end{array}}&{\begin{array}{*{20}{c}}
0\\
{\cosh g}\\
0\\
{\sinh g}
\end{array}}&{\begin{array}{*{20}{c}}
{ - \sinh g}\\
0\\
{\cosh g}\\
0
\end{array}}&{\begin{array}{*{20}{c}}
0\\
{\sinh g}\\
0\\
{\cosh g}
\end{array}}
\end{array}} \right),\tag{A8}
\end{equation}
with, ${g_1} = {g_2} = g$, ${\phi _1} = {\phi _2} = {\phi  \mathord{\left/
 {\vphantom {\phi  2}} \right.
 \kern-\nulldelimiterspace} 2}$, ${\theta _1} = 0$ and ${\theta _2} = \pi $. Hence, the propagation of $\bm{\bar X_0}$ and ${\Gamma _0}$ through SU(1,1) interferometer is determineded by $S = {S_\text{OPA2}}{S_\phi }{S_\text{OPA1}}$. And the evolution of mean values and covariance matrix of quadrature operators in SU(1,1) interferometer is given by
\begin{equation}\label{eq23}
\bm{\bar X_2} = S\bm{\bar X_0},\tag{A9}
\end{equation} 
\begin{equation}\label{eq24}
{\Gamma _2} = S{\Gamma _0}{S^T}.\tag{A10}
\end{equation} 

We obtain the mean value and covariance matrix of the outputs from the above calculation. Then, with the mean value and the covariance matrix of the lower output, we can get the measurement signal. According to Ref.\ \cite{Weedbrook2012}, the parity detection is 
\begin{equation}\label{eq44}
\langle {{{\hat \Pi }_b}} \rangle  = \frac{{\exp ( { -\bm {{\bar X}_{22}}^T \cdot {\Gamma _{22}}^{ - 1} \cdot \bm{{\bar X}_{22}}} )}}{{ \sqrt {\left| {{\Gamma _{22}}} \right|} }},\tag{A11}
\end{equation}
where, $\bm{\bar X_{22}}=( {{{\hat X}_{j,3}},{{\hat X}_{j,4}}})^T$ and ${\Gamma _{{\rm{22}}}}{\rm{ = }}\left( {\begin{array}{*{20}{c}}
{\Gamma _{\rm{2}}^{{\rm{33}}}}&{\Gamma _{\rm{2}}^{{\rm{34}}}}\\
{\Gamma _{\rm{2}}^{{\rm{43}}}}&{\Gamma _{\rm{2}}^{{\rm{44}}}}
\end{array}} \right)$.

In our scheme, the inputs are thermal and squeezed vacuum state and the initial mean value and covariance matrix of quadrature operators are 
\begin{equation}\label{eq25}
\begin{split}
&\bm{\bar X_0} = 0,
\\&{\Gamma _0} = {\sigma _\text{th}} \oplus {\sigma _\text{sqz}},
\end{split}\tag{A12}
\end{equation} 
as both of the input states have zero mean and their covariance matrices are 
\begin{equation}\label{eq50}
{\sigma _\text{th}} = \left( {2{n_{th}} + 1} \right)\left( {\begin{array}{*{20}{c}}
1&0\\
0&1
\end{array}} \right),\tag{A13}
\end{equation}
\begin{equation}\label{eq26}
{\sigma _\text{sqz}} = \left( {\begin{array}{*{20}{c}}
{{e^{2r}}}&0\\
0&{{e^{ - 2r}}}
\end{array}} \right).\tag{A14}
\end{equation}

According to Eq.\ (\ref{eq23}) and Eq.\ (\ref{eq24}), the final mean value and covariance matrix of quadrature operators become
\begin{equation}\label{eq27}
\bm{\bar X_2} = 0,\tag{A15}
\end{equation}
\begin{equation}\label{eq28}
{\Gamma _2} = \left( {\begin{array}{*{20}{c}}
{\begin{array}{*{20}{c}}
{{\gamma _{11}}}\\
{{\gamma _{21}}}\\
{{\gamma _{31}}}\\
{{\gamma _{41}}}
\end{array}}&{\begin{array}{*{20}{c}}
{{\gamma _{12}}}\\
{{\gamma _{22}}}\\
{{\gamma _{32}}}\\
{{\gamma _{42}}}
\end{array}}&{\begin{array}{*{20}{c}}
{{\gamma _{13}}}\\
{{\gamma _{23}}}\\
{{\gamma _{33}}}\\
{{\gamma _{43}}}
\end{array}}&{\begin{array}{*{20}{c}}
{{\gamma _{14}}}\\
{{\gamma _{24}}}\\
{{\gamma _{34}}}\\
{{\gamma _{44}}}
\end{array}}
\end{array}} \right),\tag{A16}
\end{equation}
where,
\begin{widetext}
\begin{equation}\label{eq29}
{\gamma _{11}} = {e^{ - 2r}}\sin ^2 ( {\frac{\phi }{2}} ){\sinh ^2}\left( {2g} \right) + \frac{1}{4}\left[ {3 + \cosh \left( {4g} \right) - 2\cos \phi {{\sinh }^2}\left( {2g} \right)} \right]\left( {1 + 2{n_\text{th}}} \right),\tag{A17}
\end{equation}
\begin{equation}\label{eq30}
{\gamma _{13}}={\gamma _{31}} = {\sin ^2}( {\frac{\phi }{2}} )\sinh \left( {4g} \right)\left( { - \cosh  r  + \sinh  r } \right)\left( {\cosh  r  + {e^r}{n_\text{th}}} \right),\tag{A18}
\end{equation}
\begin{equation}\label{eq31}
{\gamma _{14}}={\gamma _{41}} = {e^{ - 2r}}\cosh  g \sin \phi \sinh g \left[ {1 + {e^{2r}}\left( {1 + 2{n_\text{th}}} \right)} \right],\tag{A19}
\end{equation}
\begin{equation}\label{eq32}
{\gamma _{22}} = {e^{ 2r}}\sin ^2( {\frac{\phi }{2}} ){\sinh ^2}\left( {2g} \right) + \frac{1}{4}\left[ {3 + \cosh \left( {4g} \right) - 2\cos \phi {{\sinh }^2}\left( {2g} \right)} \right]\left( {1 + 2{n_\text{th}}} \right),\tag{A20}
\end{equation}
\begin{equation}\label{eq33}
{\gamma _{23}}={\gamma _{32}} = \cosh  g \sin \phi \sinh  g \left( {1 + {e^{2r}} + 2{n_\text{th}}} \right),\tag{A21}
\end{equation}
\begin{equation}\label{eq34}
{\gamma _{24}}={\gamma _{42}} = \frac{1}{2}{\sin ^2}( {\frac{\phi }{2}} )\sinh \left( {4g} \right)\left( {1 + {e^{2r}} + 2{n_\text{th}}} \right),\tag{A22}
\end{equation}
\begin{equation}\label{eq37}
{{\gamma _{33}} = \frac{1}{2}{e^{2r}}\left( {1 + \cos \phi } \right) + {e^{ - 2r}}{{\cosh }^2}\left( {2g} \right){{\sin }^2}( {\frac{\phi }{2}} ) + {{\sin }^2}\left( {\frac{\phi }{2}} \right){{\sinh }^2}\left( {2g} \right)( {1 + 2{n_\text{th}}} )},\tag{A23}
\end{equation}
\begin{equation}\label{eq38}
\begin{array}{*{20}{l}}
{{\gamma _{34}} ={\gamma _{43}} =  \cosh \left( {2g} \right)\sin  \phi  \sinh \left( {2r} \right)}
\end{array},\tag{A24}
\end{equation}
\begin{equation}\label{eq42}
{{\gamma _{44}} = \frac{1}{2}{e^{-2r}}\left( {1 + \cos \phi } \right) + {e^{  2r}}{{\cosh }^2}\left( {2g} \right){{\sin }^2}( {\frac{\phi }{2}}) + {{\sin }^2}( {\frac{\phi }{2}} ){{\sinh }^2}\left( {2g} \right)\left( {1 + 2{n_\text{th}}} \right)}. \tag{A25}
\end{equation}
\end{widetext}

\begin{widetext}
With  Eq. (\ref{eq27}) and Eq. (\ref{eq28}), we know the mean value and the covariance matrix of the lower output \emph{b} which is given as 
\end{widetext}
\begin{equation}\label{eq43}
\bm{\bar X_{22}} = \left( {\begin{array}{*{20}{c}}
0\\
0
\end{array}} \right),{\Gamma _{22}} = \left( {\begin{array}{*{20}{c}}
{{\gamma _{33}}}&{{\gamma _{34}}}\\
{{\gamma _{43}}}&{{\gamma _{44}}}
\end{array}} \right).\tag{A26}
\end{equation}
After plugging the values into Eq.\ (\ref{eq44}), we get the parity detection signal, which is given as
\begin{equation}\label{eq45}
\left\langle {{\Pi _b}} \right\rangle  = \frac{8}{{ \sqrt T }},\tag{A27}
\end{equation}
where, 
\begin{equation}\label{eq46}
\begin{split}
&T={e^{ - 2r}}\{  - 7 + 50{e^{2r}} - 7{e^{4r}} + {(1 + {e^{2r}})^2}[4\cosh (4g)\\&+3\cosh (8g) + 8\cos (2\phi ){\sinh ^4}(2g) - 8\cos \phi {\sinh ^2}(4g)]\} \\& + 32{e^{ - 2r}}{\sin ^2}\left( {\frac{\phi}{2}} \right){\sinh ^2}\left( {2g} \right){n_{th}}\{ (1 + {e^{4r}})[3 + \cosh (4g)\\&  - 2\cos \phi {\sinh ^2}(2g)] + 8{e^{2r}}{\sin ^2}(\frac{\phi}{2}){\sinh ^2}(2g)(1 + {n_{th}})\}.
\end{split}  \tag{A28}
\end{equation}

If $\phi  = 0$, the outputs will be same as the inputs and the signal will reduce to $\left\langle {{\Pi _b}} \right\rangle  = 1$ which coincides with the theory that  for one-mode squeezed vacuum, the parity signal is 1.

\section{PHASE ESTIMATION}
According to Eq. ({\ref{eq10}), Eq. ({\ref{eq44}) and Eq. ({\ref{eq43}), one can calculate the phase sensitivity via parity detection with thermal and squeezed vacuum as input states in SU(1,1) interferometer using
\begin{equation}\label{eq47}
\Delta \phi  = \frac{  {{{\Delta \hat \Pi }_b}} } {{\left| {{{\partial \langle {{{\hat \Pi }_b}} \rangle }}/{{\partial \phi }}} \right|}},\tag{B1}
\end{equation}
where,
\begin{widetext}
\begin{equation}\label{eq48}
\begin{split}
 {{\Delta \hat \Pi }_b} &=\{1-64/\{{e^{ - 2r}}\{  - 7 + 50{e^{2r}} - 7{e^{4r}} + {(1 + {e^{2r}})^2}[4\cosh (4g)+3\cosh (8g)  + 8\cos (2\phi ){\sinh ^4}(2g)\\&- 8\cos \phi {\sinh ^2}(4g)]\}   + 32{e^{ - 2r}}{\sin ^2}( {\frac{\phi}{2}}){\sinh ^2}\left( {2g} \right){n_\text{th}}\{ (1 + {e^{4r}})[3 + \cosh (4g)  - 2\cos \phi {\sinh ^2}(2g)] \\&+ 8{e^{2r}}{\sin ^2}(\frac{\phi}{2}){\sinh ^2}(2g)(1 + {n_\text{th}})\}\}\}^{1/2}, 
\end{split}\tag{B2}
\end{equation}
\begin{equation}\label{eq49}
\begin{split}
| {{{\partial \langle {{\hat \Pi _b}} \rangle } \mathord{\left/
 {\vphantom {{\partial \left\langle {{\Pi _b}} \right\rangle } {\partial \phi }}} \right.
 \kern-\nulldelimiterspace} {\partial \phi }}} |&= - \{ 128{\sinh ^2}(2g)\{  - 2\sin (2\phi ){\sinh ^2}(2g)[{\cosh ^2}r 
 + {n_\text{th}}(1 + \cosh (2r) + {n_\text{th}})]+ \sin (\phi )\{ 4{\cosh ^2}(2g){\cosh ^2}r
  \\& + 4{n_\text{th}}[{\cosh ^2}(2g)\cosh (2r) + {\sinh ^2}(2g)(1 + {n_\text{th}})]\} \} \}  /\{ {e^{ - 2r}}\{  - 7 + 50{e^{2r}} - 7{e^{4r}} + {(1 + {e^{2r}})^2}[4\cosh (4g) \\& + 3\cosh (8g) + 8\cos (2\phi ){\sinh ^4}(2g)  - 8\cos \phi {\sinh ^2}(4g)] + 32{\sin ^2}(\frac{\phi}{2}){\sinh ^2}(2g){n_\text{th}} [(1 + {e^{4r}})(3 + \cosh (4g)\\& - 2\cos \phi {\sinh ^2}(2g))+8{e^{2r}}{\sin ^2}(\frac{\phi}{2}){\sinh ^2}(2g)(1 + {n_\text{th}})]\} {\} ^{3/2}}. 
\end{split}\tag{B3}
\end{equation}

One can verify that the phase sensitivity with $\phi  = 0$ is given as in Eq.\ (\ref{eq11}).
\end{widetext}

\bibliographystyle{apsrev4-1}

\end{document}